# Discovery of ferromagnetism with large magnetic anisotropy in ZrMnP and HfMnP

Tej N. Lamichhane,[1,2] Valentin Taufour,[1,2] Morgan W. Masters,[1] David S. Parker,[3] Udhara S. Kaluarachchi,[1,2] Srinivasa Thimmaiah,[2] Sergey L. Bud'ko,[1,2] and Paul C. Canfield[1,2]
[1)] *Department of Physics and Astronomy, Iowa State University, Ames, Iowa 50011, U.S.A.*
[2)] *The Ames Laboratory, US Department of Energy, Iowa State University, Ames, Iowa 50011, USA*
[3)] *Materials Science and Technology Division, Oak Ridge National Laboratory, Oak Ridge, TN 37831, USA*

ZrMnP and HfMnP single crystals are grown by a self-flux growth technique and structural as well as temperature dependent magnetic and transport properties are studied. Both compounds have an orthorhombic crystal structure. ZrMnP and HfMnP are ferromagnetic with Curie temperatures around 370 K and 320 K respectively. The spontaneous magnetizations of ZrMnP and HfMnP are determined to be 1.9 $\mu_B$/f.u. and 2.1 $\mu_B$/f.u. respectively at 50 K. The magnetocaloric effect of ZrMnP in term of entropy change ($\Delta S$) is estimated to be $-6.7$ kJm$^{-3}$K$^{-1}$ around 369 K. The easy axis of magnetization is [100] for both compounds, with a small anisotropy relative to the [010] axis. At 50 K, the anisotropy field along the [001] axis is $\sim 4.6$ T for ZrMnP and $\sim 10$ T for HfMnP. Such large magnetic anisotropy is remarkable considering the absence of rare-earth elements in these compounds. The first principle calculation correctly predicts the magnetization and hard axis orientation for both compounds, and predicts the experimental HfMnP anisotropy field within 25 percent. More importantly, our calculations suggest that the large magnetic anisotropy comes primarily from the Mn atoms suggesting that similarly large anisotropies may be found in other 3d transition metal compounds.

In recent years, both the increase in the price of rare-earths used in magnets and adverse environmental impacts associated with their mining and purification have made the search for rare-earth-poor or rare-earth-free permanent magnets crucial. In an attempt to look for potential rare-earth-free alternatives, we studied the magnetocrystalline anisotropy of the Fe-rich compounds $(Fe_{1-x}Co_x)_2B$[1] and $Fe_5B_2P$.[2] Specifically, $(Fe_{0.7}Co_{0.3})_2B$ has drawn a lot of attention as a possible permanent magnet because of its axial magnetocrystalline anisotropy.[3–5] $Fe_5B_2P$ has a comparable magnetocrystalline anisotropy.[2]

Mn, like Fe, offers some of the highest ordered moment values, but finding Mn-based ferromagnets is challenging. Fortunately Mn is known to form ferromagnetic phases such as Mn$X$, where $X$ is a pnictogen. Recently MnBi, both in pure form and in a high temperature phase, stabilized with Rh,[6–9] has been studied as a possible Mn-based ferromagnet with moderate magnetic anisotropy.

Given the existence of Mn-$X$ ferromagnetism and our recent efforts to discover ternary ferromagnets through the surveys of transition metal - pnictogen and chalcogen ternary compounds, we used the Mn-rich, Mn-P eutectic as a starting point for a search for Mn-P-$X$ ternary ferromagnets. During our survey ($X$ = B, Al, Si, Ti, Fe, Co, Ni, Ge, Y, Zr, Nb, Rh, Pd and Hf), we discovered that ZrMnP and HfMnP are ferromagnetic at room temperature. Both ZrMnP and HfMnP have the orthorhombic crystal structure [space group: Pnma (62)].[10,11] These are bi-transition metal phosphides with TiNiSi-type structure which is an anti-$PbCl_2$ type superstructure.[11] In this paper, we report the magnetic properties (both experimental and first principle calculations) of single crystalline HfMnP and ZrMnP . We found a large magnetic anisotropy in particular for HfMnP.

The ZrMnP and HfMnP single crystals were grown by a solution growth technique[12–14] as described in Supplementary materials(SM). The structural characterization are in agreement with previous reports[10,11](see SM). The refined composition from single crystal X-ray diffraction of ZrMnP was found to be stoichiometric within two standard deviations, and HfMnP showed off-stoichiometric composition i.e., $Hf_{1.04(1)}Mn_{1.06(1)}P_{0.90(1)}$.

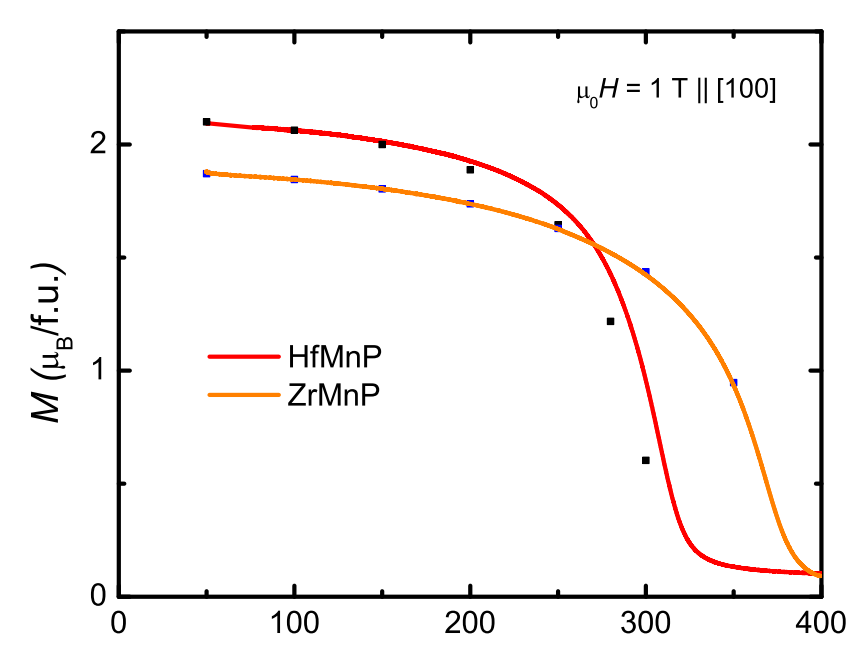

FIG. 1. Temperature dependent (solid lines) and spontaneous magnetization $M_s$ (corresponding squares) of ZrMnP and HfMnP.

Figure 1 shows the spontaneous magnetization ($M_s$) and temperature dependent magnetization of ZrMnP and HfMnP. $M_s$ was determined by the linear extrapolation from the high-field region of the easy axis [100] *M(H)* isotherm to zero field. The spontaneous magnetization of HfMnP ($2.1\mu_B$ /f.u. at 50 K) is slightly higher than ZrMnP ($1.9\mu_B$ /f.u. at 50 K). The temperature dependent magnetization is reported for the [100] direction in a 1 T applied field. Both ZrMnP and HfMnP undergo a ferromagnetic transition between 300 K and 400 K. More precise determinations of the Curie temperatures using Arrott plots (see SM) or change in slope of temperature dependent electrical resistivity (see SM) give $T_C$ around 370 K and 320 K for ZrMnP and HfMnP respectively. The anisotropy field between [100] and [010] for ZrMnP was determined to be 0.40 T at 50 K as shown in FIG. 2 (a). In HfMnP, the anisotropy field between [100] and [010] was determined to be 0.66 T at 50 K as shown in FIG. 2 (b). The anisotropy field between the hardest axis [001] and two almost degenerate axes [100] and [010] is estimated by extrapolation of tangents from the linear region of *M(H)* curves. The anisotropy field for [001] axis for ZrMnP was determined to be 4.6 T for ZrMnP and 10 T for HfMnP.

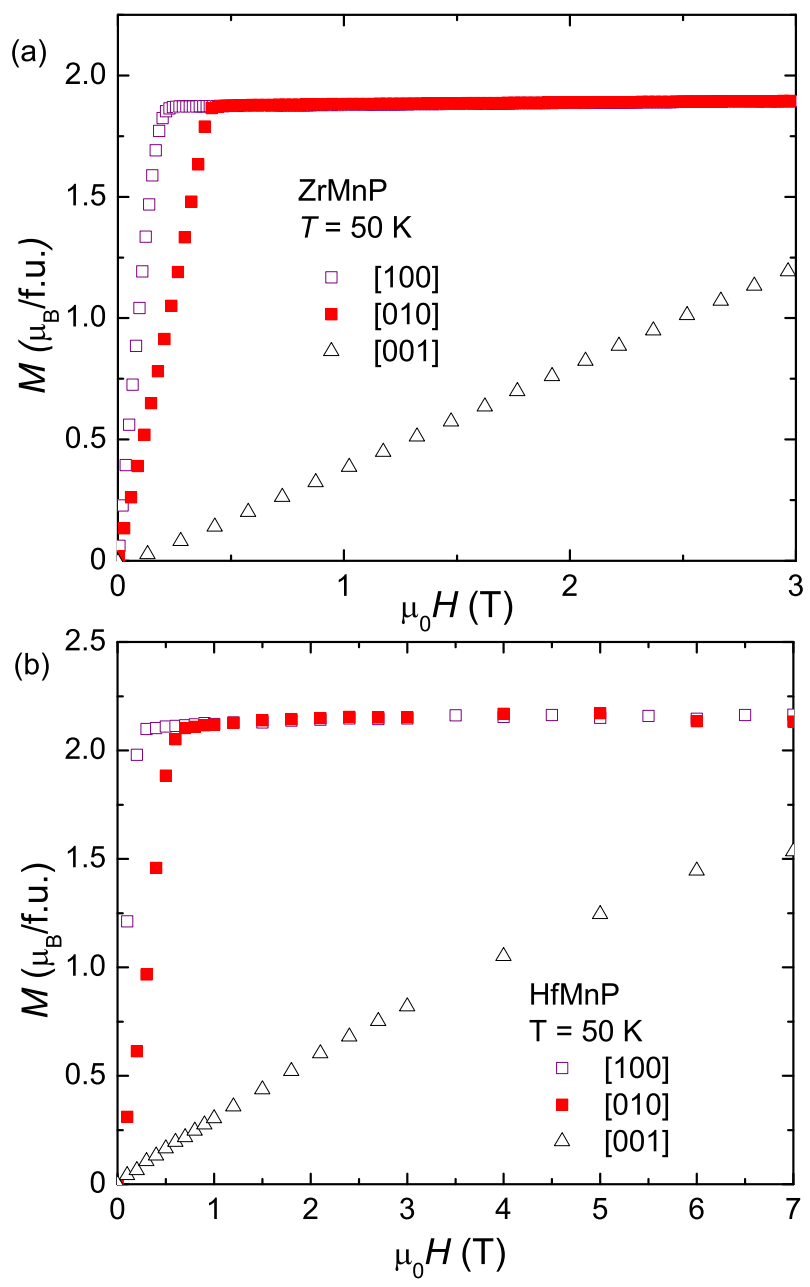

FIG. 2. Anisotropic magnetization of ZrMnP and HfMnP along salient directions at 50 K.

Comparisons for the low-temperature anisotropy of HfMnP are not abundant in the literature. $Nd_2Fe_{14}B$ has an anisotropy field around 8.2 T at room temperature.[15] Examples of reported rare-earth free permanent magnets include: $Mn_xGa$ thin-film ($H_a$ = 15 T),[16] CoPt ($H_a$ = 14 T),[17] FePt ($H_a$ = 10 T),[18] FePd ($H_a$ = 4 T at 4 K),[19] MnBi ($H_a$ = 5 T)[20] and MnAl ($H_a \approx 3$ T).[21] The HfMnP magnetocrystalline aniostropy falls in-between those values. The 10 T anisotropy of HfMnP at 50 K nearly equals with the magnetocrystalline anisotropy of $Nd_2Fe_{14}B$ at 247 K.[15]

For ZrMnP, we have studied the magnetocaloric effect from the magnetization isotherms. The results (in terms of magnetic entropy change ($\Delta S$)) are shown in FIG. 3. The relevant mathematical calculation is presented in SM. The largest effect is observed near 369 K with $-6.7$ kJ m$^{-3}$ K$^{-1}$. For comparison, the value for Gd at 294 K is $-22$ kJ m$^{-3}$ K$^{-1}$ [22 and 23], and the value of $La_{0.8}Na_{0.2}MnO_3$ at 337 K is $-10.7$ kJ m$^{-3}$ K$^{-1}$ [24]. This result indicates that ZrMnP might be a promising candidate as a magnetic refrigerant above room temperature.

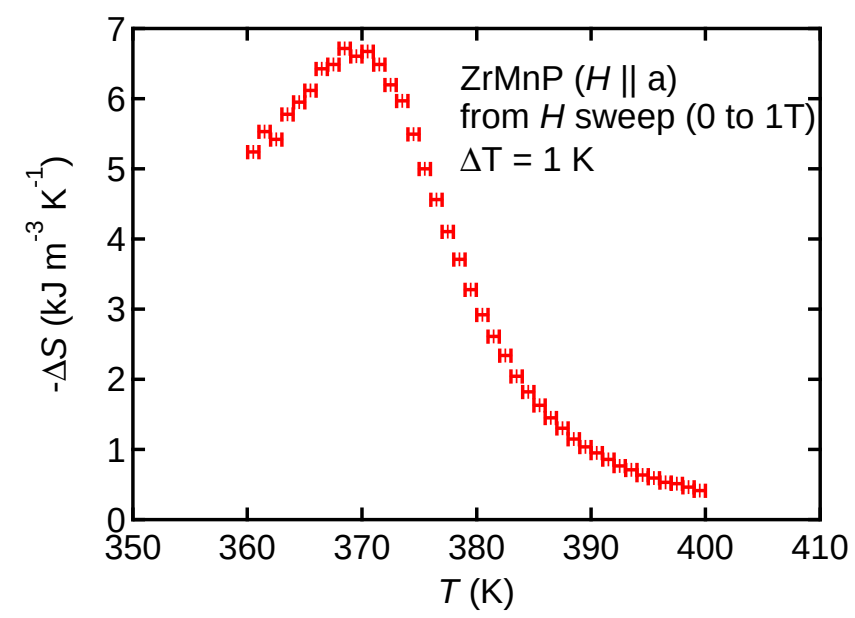

FIG. 3. Magnetic entropy change ($\Delta S$) of ZrMnP with the field change of 1 T applied along the a axis, determined from magnetization measurements.

To understand the observed magnetic behavior we have performed first principles calculations of ZrMnP and HfMnP using the plane-wave density functional theory code WIEN2K,[25] employing the generalized gradient approximation of Perdew, Burke and Ernzerhof.[26] Additional details are given in SM. For HfMnP, calculations were run for two scenarios, since the experimental results find some disorder in the sample. The first run assumes perfect stoichiometry. For the second run we have used the virtual crystal approximation (VCA) as follows. We assume P and Mn to be trivalent and Hf tetravalent, so that the 10 percent Mn occupancy of the P site is assumed isoelectronic and the 4 percent Hf occupancy on the Mn site is modeled by adding 0.04 electrons per Mn and taking the Mn ion core to have charge 25.04. Unless mentioned explicitly, the HfMnP results refer to this VCA result.

We find saturation magnetic moments of 2.02 and 1.99 $\mu_B$/f.u. for the Zr and Hf compounds, respectively. This total includes orbital moments for the Zr compound of 0.034 per Mn and 0.008 per Zr, and for HfMnP of 0.060 per Mn and 0.012 per Hf. These totals are in reasonably good agreement with the experimental values presented previously (1.9 $\mu_B$/f.u. for ZrMnP and 2.1 $\mu_B$/f.u. for HfMnP). We suspect that the slight theoretical understatement for the Hf compound is related to the non-stoichiometry of the HfMnP sample. For the Zr compound we have also estimated the Curie point by computing the energy difference (relative to the groundstate) of a configuration with nearest neighbor Mn pairs anti-aligned; one third this energy difference is taken as the Curie point, which we calculate as 654 K. This is significantly above the experimental value of around 370 K and suggests, as often occurs, that thermal fluctuations beyond the mean-field approach are important here.

Magnetic anisotropy for an orthorhombic system is characterized by total energy calculations for the magnetic moments along each of the three principal axes.[27] To first order, one can expand this energy as

$$E = E_0 + K_{aa}k_a^2 + K_{bb}k_b^2 + K_{cc}k_c^2 \quad (1)$$

where $k_{ii}$ is the magnetic moment direction cosine along axis $i$ and $K_{ii}$ the corresponding anisotropy constant.

For ZrMnP, we find, in good agreement with experiment, the [010] and [100] axes to be the 'easy' directions, separated by just 0.012 meV per Mn (in the calculation the [010] axis is the easy axis). The [001] direction is the 'hard' direction, falling some 0.136 meV per Mn above the [010] axis. On a volumetric basis this last energy difference is 0.49 MJ/m$^3$. The calculated saturation magnetic moment is 0.53 T and yields a calculated anisotropy field of 2.3 T. This is somewhat less than the 4.6 T anisotropy field observed in the experiment. One recalls a similar discrepancy in the MnBi ferromagnet,[8,28,29] which has been argued as arising from correlation or lattice dynamics effects; it is possible that similar effects are at work here.

For HfMnP (in the VCA calculation) we find, again in good agreement with experiment, the [010] and [100] axes as the 'easy' directions, separated by 0.041 meV/Mn; as with ZrMnP the [010] axis is calculated to be the easy axis. The [001] direction lies some 0.47 meV per Mn above the [010] direction. This is a much larger anisotropy than found in the Zr compound and could suggest the importance of the Hf atoms in the anisotropy. To check this we have run a calculation in which spin-orbit coupling (the source of magnetocrystalline anisotropy) is turned off for the Mn atom, and computed the anisotropy. In fact we find from this calculation that just 30 percent of the magnetic anisotropy arises from the Hf atom - fully 70 percent arises from the Mn atom. This is a surprising result given that it is widely assumed that heavy elements (such as the rare earth elements) are indispensable for magnetic anisotropy. Here it is in fact the 3*d* element Mn that generates most of the anisotropy.

On a volumetric basis the total anisotropy is 1.78 MJ/m$^3$, yielding an anisotropy field of 8.1 T which is comparable with the 10 T value from the experiment. The discrepancy could arise from the effects mentioned for ZrMnP, or perhaps from the disorder in the sample. It is worth noting that our calculation of disorder-free HfMnP finds a significantly smaller anisotropy of 1.43 MJ/m$^3$. This suggests the importance of disorder to the anisotropy, and surprisingly suggests that additional disorder might in fact *increase* the anisotropy from the substantial values already found.

It is of interest to understand this large magnetic anisotropy. Presented in FIG. 4 is the calculated VCA bandstructure of HfMnP in the orthorhombic Brillouin zone in the ferromagnetic state, with spin-orbit coupling included (this mixes spin-up and spin-down states so that the bands presented contain both characters). Two facts are immediately evident from the plot; firstly, there are a large number of Fermi-level crossings, despite the general transfer of spectral weight away from the Fermi level associated with the magnetic transition. Secondly, and more importantly, there are a number of band crossings (indicated by the red ovals) that fall virtually *at* the Fermi level.

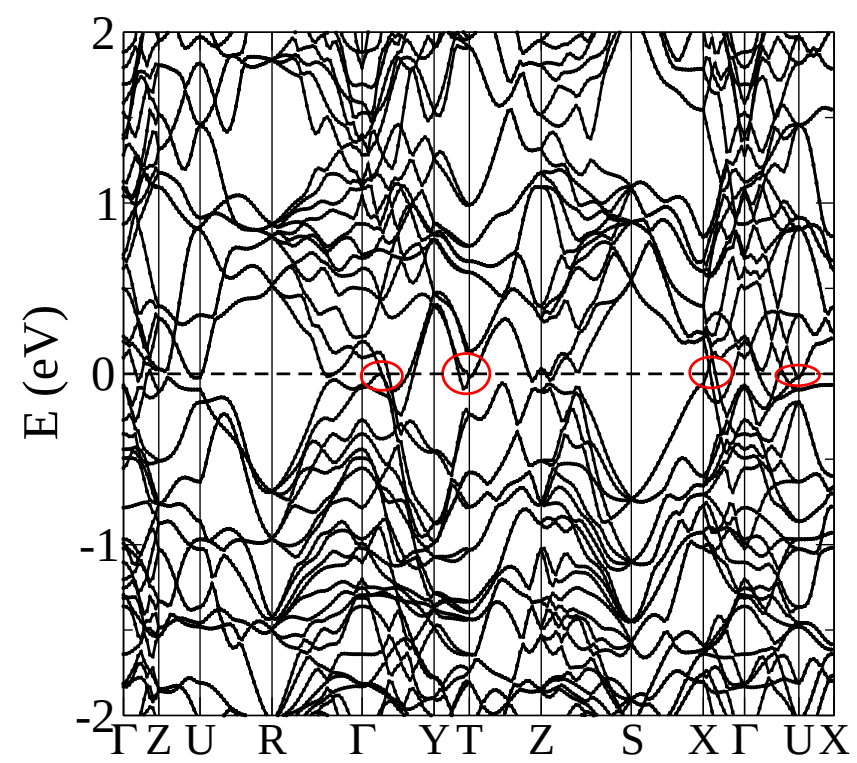

FIG. 4. The calculated bandstructure of HfMnP; note the multiple band crossings near $E_F$ indicated by red ovals.

Magnetic anisotropy arises from spin-orbit coupling and involves a sum of matrix elements connecting occupied and unoccupied states *at the same wavevector*, divided by the energy difference of these states (see Ref. 30 for the exact expression). From this fact it is clear that magnetic materials that have a large amount of band crossings very near $E_F$ will generally have a larger magnetic anisotropy, since this allows the coupling of occupied and unoccupied states with the same wavevector and small energy denominator. HfMnP has several such features and it is most likely these that cause the large magnetic anisotropy. We note also that the magnetic anisotropy effect of these crossings will be sensitive to their exact position relative to $E_F$ and thus it is plausible that disorder effects on the anisotropy are substantial, as we observe.

Perhaps the most important finding from the theoretical part of the work is the large magnetic anisotropy - 0.47 meV per Mn - found for HfMnP. On a per atom basis this is nearly *eight* times the experimental value of 0.06 meV per atom for hcp Cobalt.[31] It is probable that this large anisotropy is not specific to Mn, but originates in the unusual electronic and anisotropic orthorhombic structure. If it were possible to stabilize a 3*d*-based ferromagnet with this anisotropy, but a higher proportion of the 3*d* element, one might obtain a very useful magnetic material without the need for expensive rare-earth elements. Since materials with high proportions of Mn tend towards antiferromagnetism, rather than the desired fer-romagnetic alignment, such a magnetic material might best be based on Fe, which often aligns ferromagnetically. One example of such a material is $Fe_2P$, which shows an MAE of 2.32 $MJ/m^3$[30], although this material has a low Curie temperature. In summary, HfMnP and ZrMnP single crysals were grown by a self-flux growth technique and their structural, transport and magnetic properties were studied. Both of the phosphides are ferromagnetic at room temperature. The Curie temperatures of ZrMnP and HfMnP are around 370 K and 320 K respectively. The easy axis of magnetization is [100] for both compounds with a small anisotropy field along [010] axes. The anisotropy field along [001] axes is found to be 4.5 T and 10 T for ZrMnP and HfMnP respectively. The spontaneous magnetizations of ZrMnP and HfMnP at 50 K are determined to be $1.9\mu_B$/f.u. and $2.1\mu_B$/f.u. respectively. The magnetic entropy change ($\Delta S$) of ZrMnP is found to be $-6.7$ $kJm^{-3}K^{-1}$ near its Curie temperature. The observed magnetic properties are well-explained with the first principle calculation results.

## SUPPLEMENTARY MATERIALS

See supplemental materials at [url will be inserted by aip] for details about crystal growth, structural characterization, crystallographic orientation, resistivity measurements, magnetization measurements, magnetocaloric analysis and additional details of first-principle calculation.

## ACKNOWLEDGEMENT

We would like to thank T. Kong, G. Drachuck, W. Meier for useful discussions. Dr. Warren Straszheim is acknowledged for doing WDS on various samples. TNL, VT, DSP and PCC were supported by the Critical Materials Institute, an Energy Innovation Hub funded by the U.S. Department of Energy, Office of Energy Efficiency and Renewable Energy, Advanced Manufacturing Office. ST, MWM, USK and SLB were supported by the Office of Basic Energy Sciences, Materials Sciences Division, U.S. DOE. This work was performed at the Ames Laboratory, operated for DOE by Iowa State University under Contract No. DE-AC02-07CH11358.

# Supplementary materials for "Discovery of ferromagnetism with large magnetic anisotropy in ZrMnP and HfMnP"

Tej N. Lamichhane,[1,2] Valentin Taufour,[1,2] Morgan W. Masters,[1] David S. Parker,[3] Udhara S. Kaluarachchi,[1,2] Srinivasa Thimmaiah,[2] Sergey L. Bud'ko,[1,2] and Paul C. Canfield[1,2]
[1)] *Department of Physics and Astronomy, Iowa State University, Ames, Iowa 50011, U.S.A.*
[2)] *The Ames Laboratory, US Department of Energy, Iowa State University, Ames, Iowa 50011, USA*
[3)] *Materials Science and Technology Division, Oak Ridge National Laboratory, Oak Ridge, TN 37831, USA*

## I. CRYSTAL GROWTH

### A. Initial test

ZrMnP and HfMnP single crystals were grown by a solution growth technique using the Mn-rich Mn-P eutectic composition ($Mn_{87}P_{13}$) as a starting solution.[1] Establishing the safe handling of the Mn-P solution for the crystal growth process was carried out in a manner similar to our previous phosphorous-based works.[2,3] Freshly ball-milled Mn powder (Alfa Aesar, 99.9%) and red P lumps (Alfa Aesar, 99.999%) were mixed in the eutectic composition in a fritted alumina crucible set[4,5] and sealed in an amorphous silica tube under a partial pressure of Ar. The Mn powder was covered with red P lumps in the growth crucible. Then the ampoule was heated up to 250 °C over 3 h and kept at 250 °C for 3 h. These steps are taken to avoid the vaporization of P by immediately mixing the molten P with Mn powder. The ampoule was then slowly heated to 1150 °C over 12 h and held there for 3 h to better homogenize the melt. Finally, the ampoule was cooled down to 1000 °C over 50 h and decanted with the help of a centrifuge. All the content was recovered in the catch crucible clearly demonstrating that all material forms a liquid at 1000°C and above. Given that there was no evidence of significant P-vapour pressure, this confirmed the safe and controlled handling of Mn-P solution for single crystal growth.

### B. Mn cleaning

Elemental Mn tends to oxidize, and a surface layer of MnO forms on exposure to the atmosphere. If the Mn used contains a significant amount of MnO, a shift of the exact stoichiometry can result in growth of $Mn_3P$ crystals in the above-mentioned eutectic test. To reduce the MnO content of the Mn used, Mn plates were heat-treated inside an amorphous $SiO_2$ tube under a partial Ar atmosphere. The mechanism of Mn cleaning is the following:[6]

$$2\mathrm{MnO} + \mathrm{SiO_2} \xrightarrow{950^\circ\mathrm{C}\ (\text{Up to } 75\ \mathrm{h})} \mathrm{Mn_2SiO_4}$$

with the resulting $Mn_2SiO_4$ deposited on the interior wall of the tubing.

### C. Single Crystal Growth

The cleaned Mn plates were ball-milled using an agate ball-milling set for 2 minutes. In the ternary crystal growth attempts, various amounts of elemental metal pieces (Zr or Hf) were covered with Mn powder and P lumps, in the same fashion as in our eutectic test. To make sure that the second transition metal was dissolved in the melt, the growth ampoules were heated up to 1180 ° C and held there for 3 h. The growth ampoules were then cooled down to 1025 ° C over 180 h where the flux was decanted via centrifuge. By careful observation of crystals sizes, 1.25% of Zr or Hf metal in the Mn-P eutectic was found to be the optimal stoichiometry ($Zr_{1.25}Mn_{85.90}P_{12.85}$ and $Hf_{1.25}Mn_{85.90}P_{12.85}$) to get larger (but still relatively small) crystals. The crystals were partially covered with flux and etched with 1 molar HCl solution. The acid-etched needle like single crystals of ZrMnP and HfMnP are shown in FIG. S 1(a) and 2(a) respectively.

## II. STRUCTURAL CHARACTERIZATION

The single crystal XRD data were obtained using a Bruker APEX2 diffractometer. Single crystals of ZrMnP and HfMnP were mounted on a goniometer head of a Bruker Apex II CCD diffractometer and measured using graphite-monochromatized $MoK_\alpha$ radiation ($\lambda = 0.71073 A^\circ$). Reflections were gathered at room temperature by taking four sets of 360 frames with 0.5° scans in $\omega$, with an exposure time of 20 s (ZrMnP) or 30 s (HfMnP) per frame. The crystal-to-detector distance was 50 mm. The reflections were collected over the range of $4.2 < \theta < 31.3°$ and corrected for Lorentz and polarization effects. The intensities were further corrected for absorption using the program SADABS, as implemented in Apex 2 package.[7]

The structure solution and refinement of ZrMnP and HfMnP were carried out using SHELXTL[8] in the space group Pnma.[9,10] The final stage of refinement was performed using anisotropic displacement parameters for all the atoms. The refined composition of ZrMnP was found to be stoichiometric within two standard deviations, whereas HfMnP showed off-stoichiometric composition i.e., $Hf_{1.04(1)}Mn_{1.06(1)}P_{0.90(1)}$. Tables S I, S II and S III summarize data collection, lattice parameters, atomic positions, site occupancy factors, and displacement parameters for crystals of ZrMnP and HfMnP.

To get the powder XRD data, several etched single crystalline needles of each phosphide were finely powdered and spread over a zero background single crystalline silicon wafer with help of a thin film of Dow Corning high vacuum grease and scanned using Rigaku Miniflex X-ray diffractometer with Cu $K_\alpha$ radiation source. Scans were performed over 90° in 0.01° increments and data was acquired over 5 sec exposures. The atomic co-ordinate information from the single crystal measurements were used to refine the powder XRD patterns (Rietveld analyses) of ZrMnP and HfMnP samples. The $R_p$ for both Rietveld analyses were less than 0.081. The Rietveld analysed powder XRD patterns for ZrMnP and HfMnP are shown in FIG. S 1(e) and 2(e) respectively.

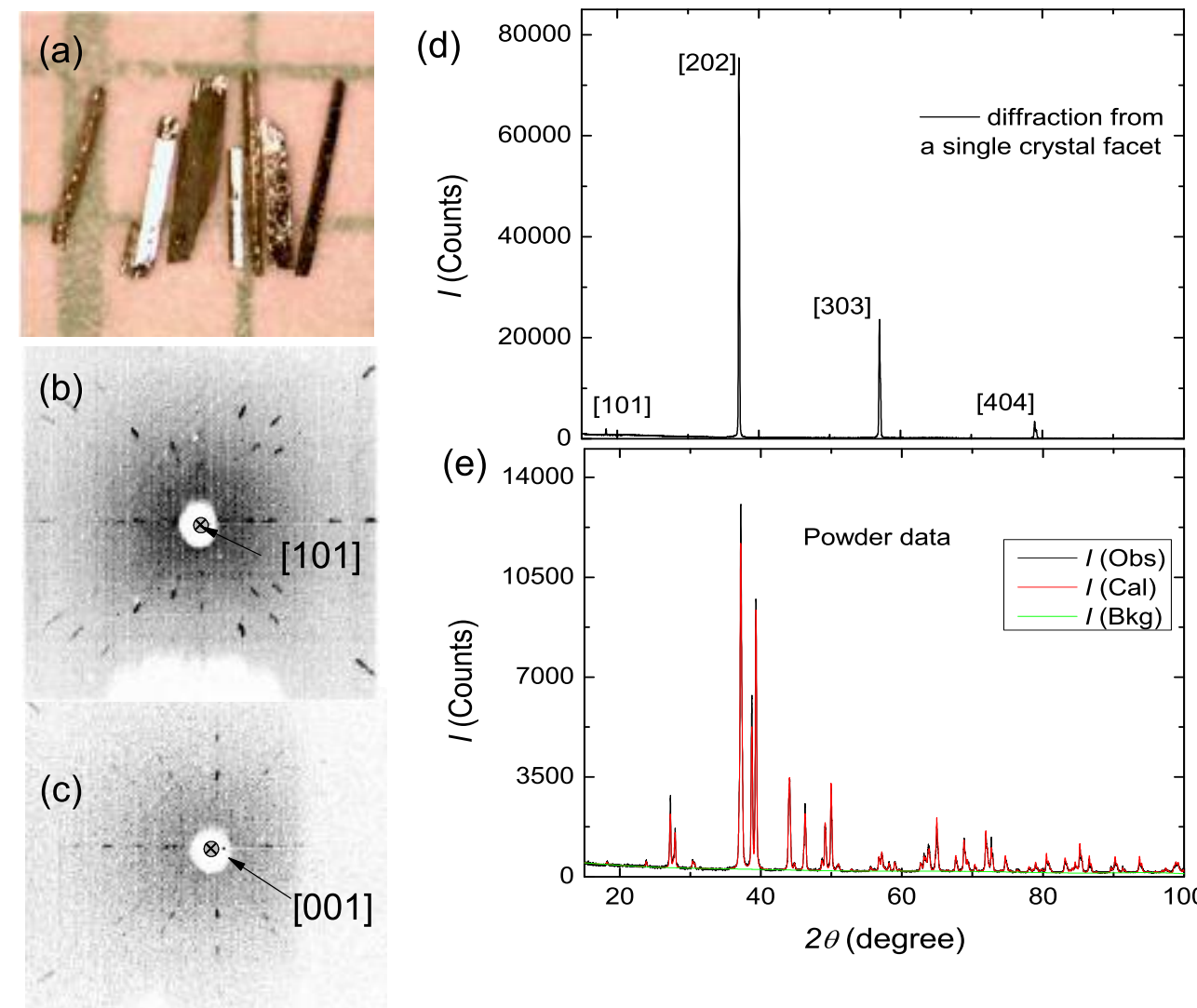

FIG. S 1. (a) Single crystals of ZrMnP (b) Back-scattered Laue pattern from facet [101] (c) Back-scattered Laue pattern [001] which was obtained at an angle of ≈ 49° in an anti-clockwise direction to [101] (d) X-ray diffraction from a single crystal facet keeping the Rigaku Miniflex XRD puck fixed (e) Rietveld refined powder XRD data.

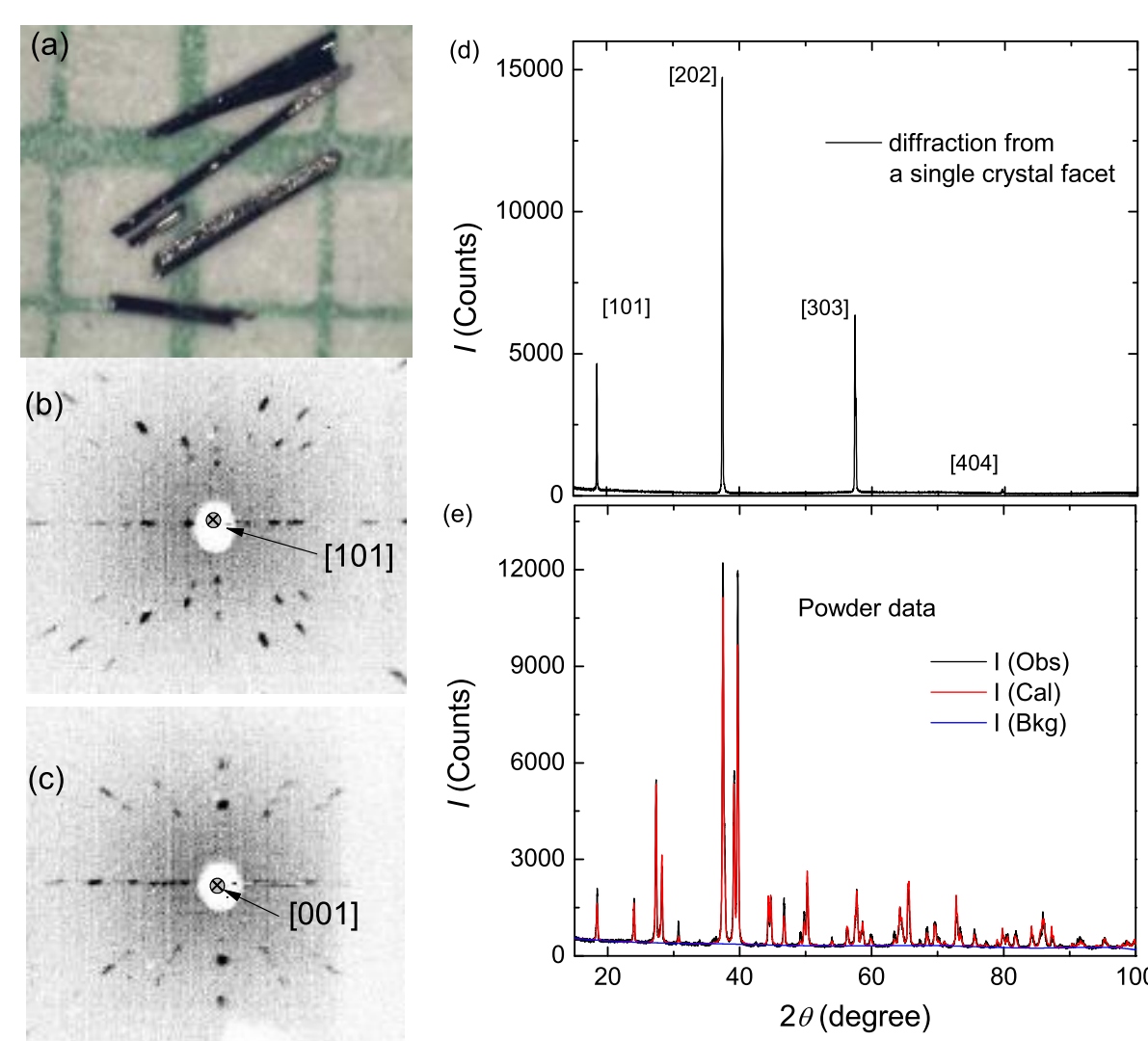

FIG. S 2. (a) Single crystals of HfMnP (b) Back-scattered Laue pattern from a facet [101] (c) Back-scattered Laue pattern pattern [001] which was obtained at an angle of ≈ 49° in an anti-clockwise direction to [101] (d) X-ray diffraction from a single crystal facet keeping the Rigaku Miniflex XRD puck fixed (e) Reitveld refined powder XRD data.

## III. CRYSTALLOGRAPHIC ORIENTATION

The crystallographic orientations of single crystalline samples were determined by analysing backscattered Laue photographs using the OrientExpress software package[11] and the diffraction of monochromatic X-ray beam from the sample facets over the Bragg Brentano Geometry with the beam direction parallel to the scattering vector . The normal to the flat facets of the as-grown rods is [101] or equivalent direction (see Laue photographs of ZrMnP and HfMnP sample in of the FIG. S 1(b) and 2(b)) and the long axis of the rod is [010]. To confirm the normal to the flat faces is [101] or [−101], X-ray diffraction peaks from single crystal facets of ZrMnP and HfMnP were obtained from the powder XRD machine (Bragg Brentano Geometry on Rigaku Miniflex diffractrometer). The X-ray diffraction on the single crystal facets gave the four diffraction peaks of $\{i0i\}$ type as shown in FIG. S 1(d) and 2(d). Then the samples facing [101] direction toward the X-Ray beam were rotated 49° in anti-clockwise direction with the help of a goniometer to get [001] Laue patterns as shown in FIG. S 1(c) and 2(c).

## IV. RESISTIVITY MEASUREMENTS

The ZrMnP and HfMnP needles were suitable for resistivity measurement. The four probe measurement technique was employed to measure the resistivity. Electrical contacts were prepared by gluing platinum wires with H20E EPO-TEK epoxy. Quantum Design (QD) Magnetic Property Measurement System (MPMS) was used as a temperature environment during the resistivity measurement. The MPMS was interfaced with a Linear Research, Inc. ac resistance bridge (LR 700).

Table S I. Crystal data and structure refinement for ZrMnP and HfMnP.

| | | |
|---|---|---|
| Empirical formula | ZrMnP | $Hf_{1.04(1)}$ $Mn_{1.06(1)}$ $P_{0.90(1)}$ |
| Formula weight | 177.13 | 271.37 |
| Temperature | 293(2) K | 293(2) K |
| Wavelength | 0.71073 Å | 0.71073 Å |
| Crystal system, space group | Orthorhombic, Pnma | Orthorhombic, Pnma |
| Unit cell dimensions | a=6.4170(17) Å | a=6.3257(17) Å |
| | b = 3.6525(10) Å | b = 3.6298(10) Å |
| | c = 7.5058(19) Å | c = 7.409(2) Å |
| Volume | 175.92(8) $Å^3$ | 170.13(8) $Å^3$ |
| Z, Calculated density | 4, 6.688 g/$cm^3$ | 4, 10.595 g/$cm^3$ |
| Absorption coefficient | 13.415 $mm^{-1}$ | 71.333 $mm^{-1}$ |
| F(000) | 320 | 459 |
| Crystal size | 0.06 x 0.04 x 0.02 $mm^3$ | 0.05 x 0.07 x 0.02 $mm^3$ |
| $\theta$ range ($^\circ$) | 4.178 to 31.307 | 4.236 to 31.405 |
| Limiting indices | $-8 \leq h \leq 9$ | $-8 \leq h \leq 8$ |
| | $-5 \leq k \leq 5$ | $-5 \leq k \leq 5$ |
| | $-10 \leq l \leq 10$ | $-10 \leq l \leq 10$ |
| Reflections collected | 2507 | 2443 |
| Independent reflections | 317 [R(int) = 0.0216] | 305 [R(int) = 0.0673] |
| Completeness to $\theta = 25.242^\circ$ | 100.00% | 100.00% |
| Absorption correction | multi-scan, empirical | multi-scan, empirical |
| Refinement method | Full-matrix least-squares on $F^2$ | Full-matrix least-squares on $F^2$ |
| Data / restraints / parameters | 317 / 0 / 20 | 305 / 0 / 22 |
| Goodness-of-fit on $F^2$ | 1.216 | 1.080 |
| Final R indices [I> $2\sigma$(I)] | $R1 = 0.0135$, $wR2 = 0.0286$ | $R1 = 0.0247$, $wR2 = 0.0460$ |
| R indices (all data) | $R1 = 0.0144$, $wR2 = 0.0291$ | $R1 = 0.0359$, $wR2 = 0.0494$ |
| Extinction coefficient | 0.0376(17) | 0.0088(2) |
| Largest diff. peak and hole | 0.542 and -0.756 e.$Å^{-3}$ | 2.381 and -2.526 e.$Å^{-3}$ |

Table S II. Atomic coordinates and equivalent isotropic displacement parameters ($A^2$) for ZrMnP. U(eq) is defined as one third of the trace of the orthogonalized $U_{ij}$ tensor.

| atom | Occ. | x | y | z | $U_{eq}$ |
|---|---|---|---|---|---|
| Zr1 | 1 | 0.0310(1) | 0.2500 | 0.6746(1) | 0.004(1) |
| Mn2 | 1 | 0.1373(1) | 0.2500 | 0.0592(1) | 0.004(1) |
| P3 | 1 | 0.2670(1) | 0.2500 | 0.3708(1) | 0.004(1) |

Table S III. Atomic coordinates and equivalent isotropic displacement parameters ($A^2$) for $Hf_{1.04}$ $Mn_{1.06}$ $P_{0.90}$. U(eq) is defined as one third of the trace of the orthogonalized $U_{ij}$ tensor.

| atom | Occ. | x | y | z | $U_{eq}$ |
|---|---|---|---|---|---|
| Hf1 | 1 | 0.0314(1) | 0.2500 | 0.6741(1) | 0.003(1) |
| Mn2 | 0.96(1) | 0.1400(1) | 0.2500 | 0.0593(1) | 0.005(1) |
| Hf2 | 0.04(1) | | | | |
| P3 | 0.90(1) | 0.2670(2) | 0.2500 | 0.3697(2) | 0.004(1) |
| Mn3 | 0.10(1) | | | | |

The resistivity of both compounds was measured along the [010] direction. The resistivity data for ZrMnP and HfMnP are shown in FIG. S 3. The resistivity of either compound is fairly metallic in nature. The residual resistivity ratio (RRR=$\frac{\rho(300K)}{\rho(2K)}$) was determined to be 5.6 and 5.4 for ZrMnP and HfMnP respectively. Although the RRR values for both of the compounds are comparable, HfMnP has a residual resistivity, $\rho_0$, that is over 50% higher, consistent with a slightly off-stoichiometric composition.

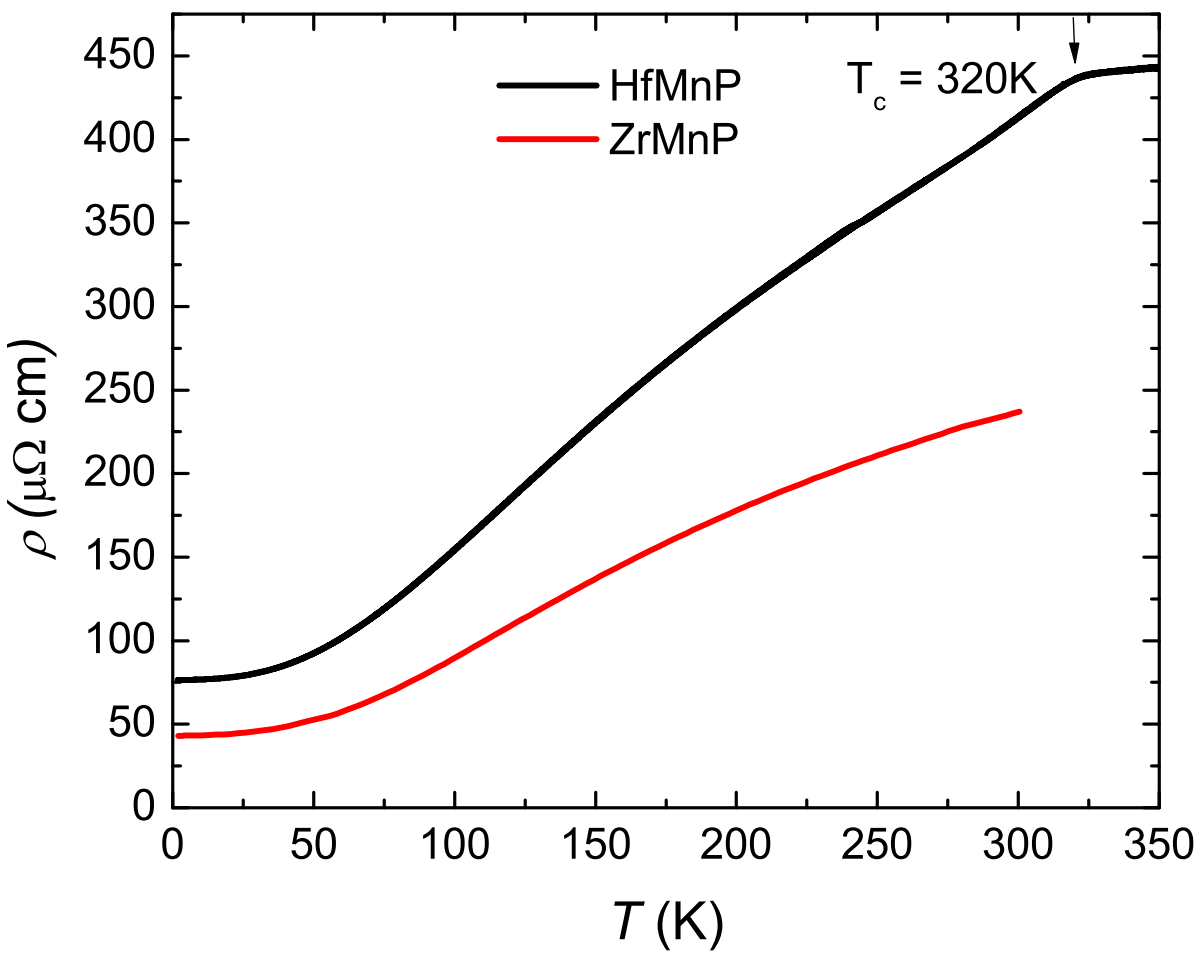

FIG. S 3. Resistivity of ZrMnP and HfMnP with an excitation current along [010] direction

In the case of HfMnP, a change in slope of resistivity occurs at ∼ 320 K, which is consistent with a ferromagnetic phase transition at that temperature. This can be seen more clearly in FIG. S 4(a) where a loss of spin-disorder-scattering feature is clearly found for $T_C \sim 320$ K. A similar value can be inferred from the low-field (0.1 T magnetization data shown in FIG. S 4(b) (The thermal hysteresis is due to heating/cooling at 3 K/minute). Based on these data, $T_C$ is determined to be around 320 K for HfMnP.

## V. MAGNETIZATION MEASUREMENTS

The magnetic properties of the samples were measured using QD Versalab Vibrating Sample Magnetometer (VSM) in standard option from 50 K to 400 K. We used GE 7031 Varnish to glue the sample on the VSM quartz stick. The etched, rod-like samples were used for magnetization measurements. Both the temperature-dependent magnetization *M(T)* (at 1 T) and field-dependent magnetization *M(H)* data (starting at 50 K in steps of 50 K) were measured for each sample. The sample glued with GE varnish was further secured with teflon tape.

### A. Identification of easy axis of magnetization and demagnetization factor

Experimental identification of the easy axis of magnetization for a needle-like crystal is not straightforward because of demagnetization effects, particularly if it is not along the axial direction. In addition, the small sample size makes the alignment difficult. The strongest criteria for the determination of the easy axis of magnetization are: (i) the highest saturation magnetization at low applied field and (ii) overlapping of Arrott plot curves at low-temperature isotherms. None of the faceted [101] or axial [010] directions of the ZrMnP and HfMnP crystals were found to be the easy axis of magnetization. So to align the [100] or [001] axis in the field direction, a delrin sample holder with 45° slope was prepared so that each facet is aligned with an error of ±4°. The magnetization of the delrin sample holder was subtracted from the measurements.

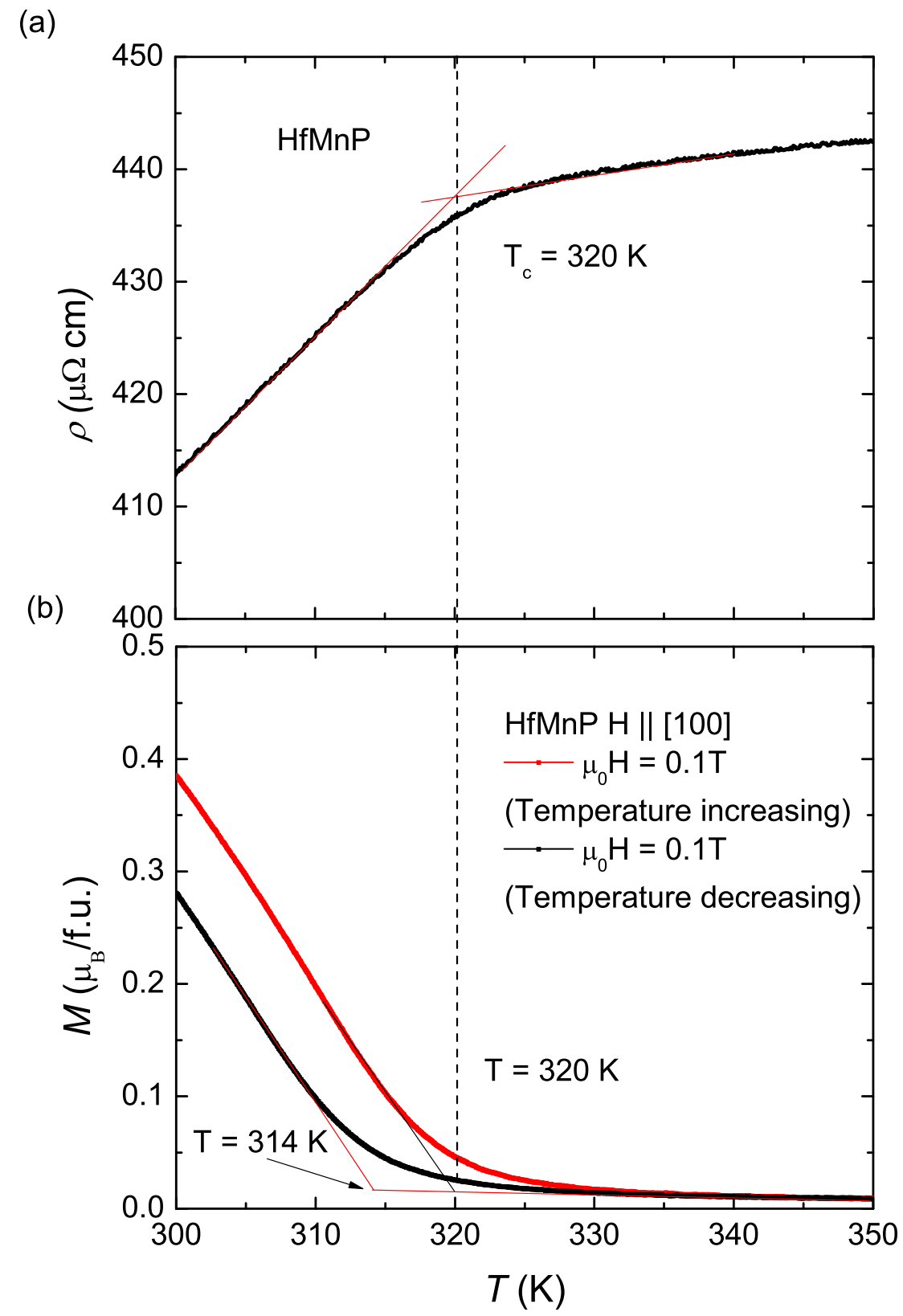

FIG. S 4. Curie Temperature determination for HfMnP. (a) Enlarged resistivity data near the anomaly in resistivity (b) Enlarged temperature dependent magnetization data near the ferromagnetic transition. The temperature of slope change in resistivity data ($T_C = 320$ K) equals to the upper limit of the temperature range of the slope change in *M(T)* data along [100] direction for 0.1 T field.

Arrott plots for all 3 principle directions of the orthorhombic rods for ZrMnP sample are shown in FIG. S 5. The detailed explanation of Arrott plot analysis for field along the easy axis of magnetization is in our previous work.[2] The Arrott curves at low temperatures along [100] direction overlap for ZrMnP, indicating this as the easy axis of magnetization. Between the two hard

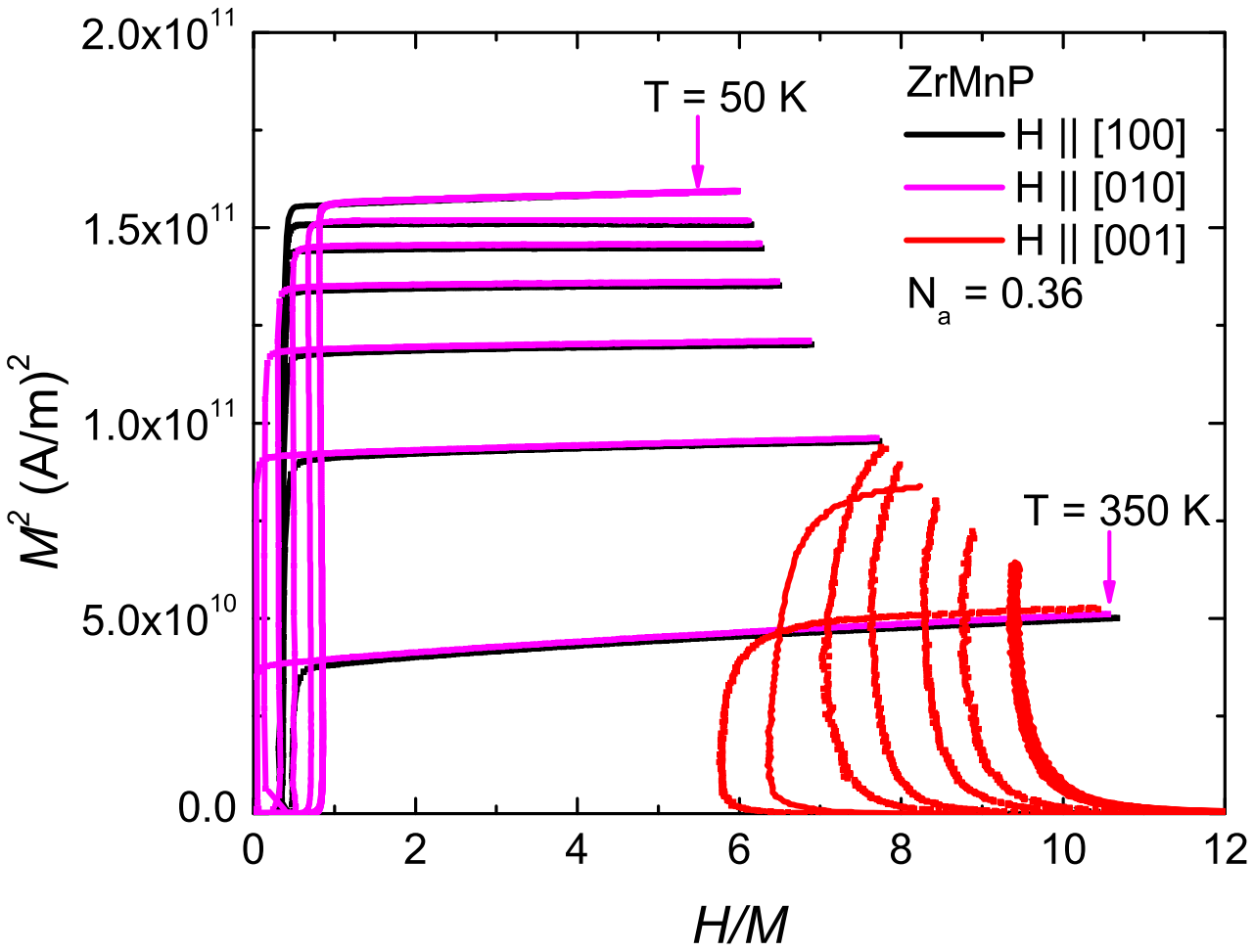

FIG. S 5. Determination of demagnetization factor for the easy axis of magnetization of ZrMnP sample: The highest value of $M^2$ curve corresponds to 50 K and lowest valued $M^2$ curve corresponds to 350 K, all others taken at intervals of 50 K for all directions measured. The overlapping of $M^2$ curves with minimum intercept in $\frac{H}{M}$ axis at low-temperature isotherms is considered to occur along the easy axis (here [100]) of the magnetization. The demagnetization factor for the [100] direction was determined to be $N_a = 0.36$.

axes, [010] is an easier direction of magnetization than [001].

To determine the anisotropy axes and fields for HfMnP at 50 K, more sensitive and higher field (7 T) QD MPMS equipped with angular magnetization measurement was used. FIG. S 6 shows the variation of magnetization of HfMnP sample with respect to the rotation angle. Starting from the [101] facet, it reaches to its minimum when [001] direction aligns with field and maximum when [100] direction aligns, consistent with the results obtained from the Arrott plot measurements in ZrMnP.

### B. Curie temperature determination

Arrot plot analyses were made for the easy axis (along [100]) of magnetization of ZrMnP to determine the Curie temperature. In an Arrott plot, $M^2$ isotherms are plotted as a function of $\frac{H_{int}}{M}$ where $H_{int} = H$-N*$M$ is the internal field of the magnetization sample after the correction of the demagnetization effect. N is the demagnetization factor which can be experimentally determined for the easy axis of magnetization.[2] $N_a$ was determined to be 0.36 for ZrMnP (see FIG. S 5). In the mean field approximation, the Arrott curve at the Curie temperature is a straight line passing through origin. The Curie temperature of ZrMnP was determined to be $\sim 370$ K as shown in FIG. S 7.

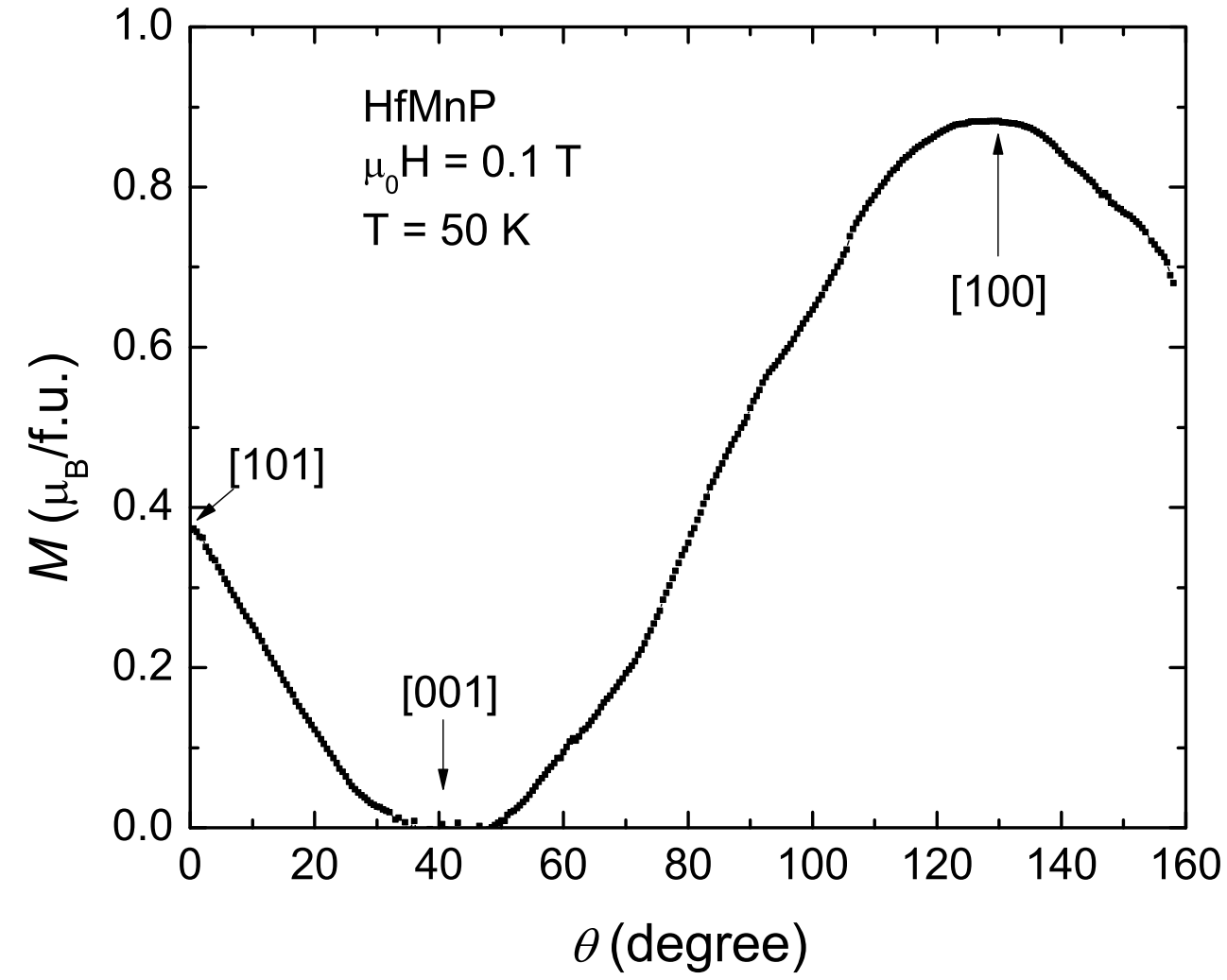

FIG. S 6. Angular magnetization of HfMnP measured at 50 K in a 0.1 T applid field.

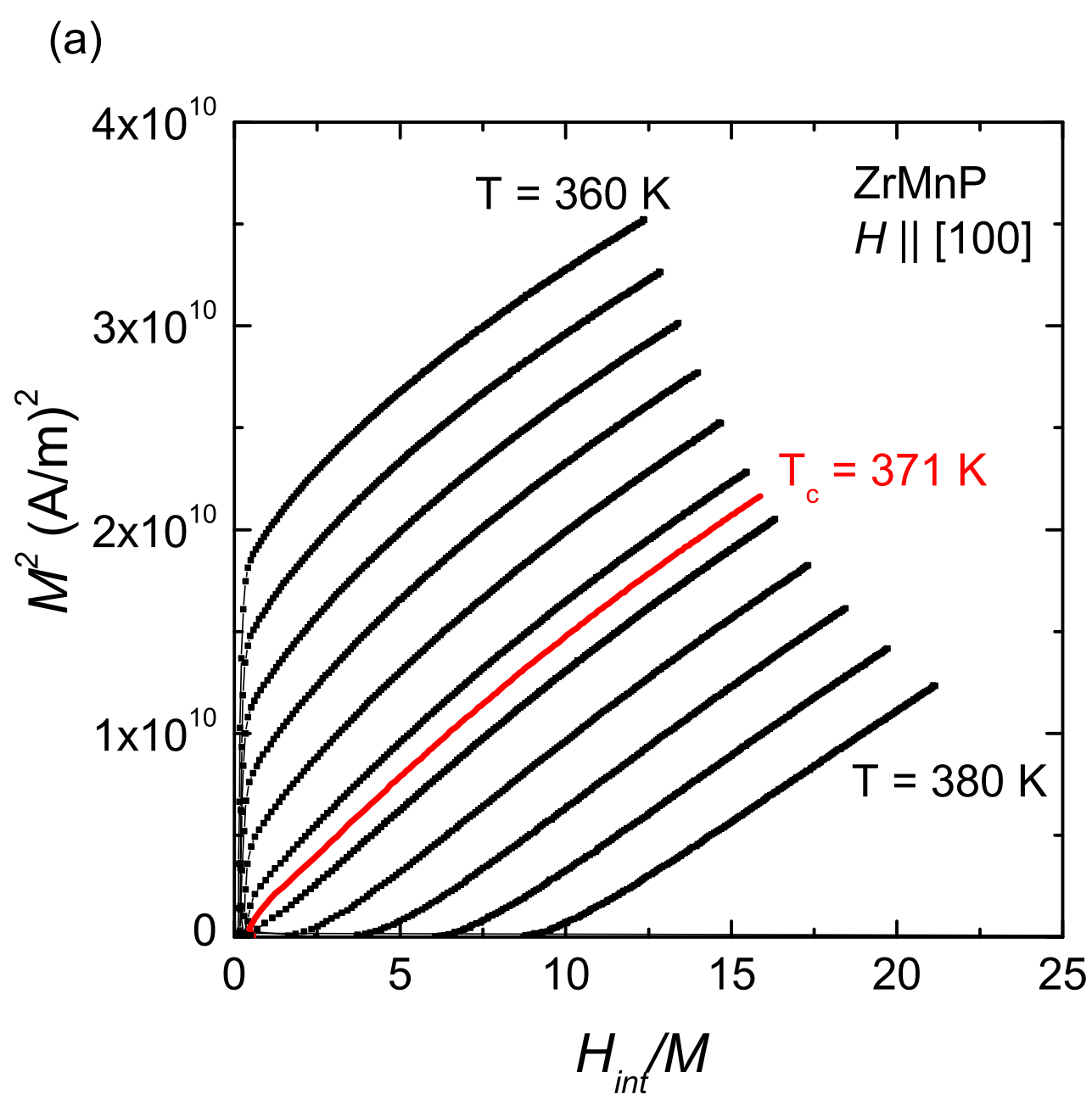

FIG. S 7. Determination of Curie temperature of ZrMnP from *M(H)* data along [100]. All black colored Arrott curvers are taken at a step of 2 K temperature.

### C. Determination of the magneto-caloric effect

The isothermal entropy change can be determined from magnetization measurements using the following equations:

$$\Delta S_{\rm iso}(T, \Delta H) = \int_{H_i}^{H_f} dS = \int_{H_i}^{H_f} \frac{\partial S}{\partial H}\bigg|_T dH$$

Using the Maxwell relation $\left.\frac{\partial S}{\partial H}\right|_T = \mu_0 \left.\frac{\partial M}{\partial T}\right|_H$, we have:

$$\Delta S_{\rm iso}(T, \Delta H) = \mu_0 \int_{H_i}^{H_f} \left.\frac{\partial M}{\partial T}\right|_H dH$$

For small enough temperature steps at $T_u$ and $T_l$ around $T$, we can approximate:

$$\left.\frac{\partial M}{\partial T}\right|_H \approx \frac{M(T_u, H) - M(T_l, H)}{T_u - T_l}$$

So that we have:

$$\Delta S_{\rm iso}(T, \Delta H) \approx \frac{\mu_0}{T_u - T_l} \int_{H_i}^{H_f} M(T_u, H) - M(T_l, H) dH \qquad \text{(S 1)}$$

The magnetization isotherms near $T_{\rm C}$ as shown in FIG. S 7 can be used with Eq. S 1 to determine $\Delta S_{\rm iso}(T, \Delta H)$.

## VI. ADDITIONAL DETAILS OF FIRST-PRINCIPLES CALCULATIONS

Calculations were performed using the experimental lattice parameters and all internal coordinates not dictated by symmetry were relaxed until internal forces were less than 2 mRyd/Bohr. Sphere radii (in Bohr) for ZrMnP of 1.86 for P, 2.36 for Mn and 2.5 for Zr were used; for HfMnP sphere radii of 1.97 for P, 2.3 for Mn and 2.5 for Hf were used. An $RK_{max}$ (the product of the smallest sphere radius, i.e. P, and the largest planewave expansion wavevector) of 7.0 was used and the linearized augmented plane-wave basis was used throughout. In general all calculations used sufficient numbers of $k$-points - a minimum of $1,000$ $k$ points in the full Brillouin zone for convergence, and as many as $30,000$ $k$-points for the calculations of magnetic anisotropy. For this anisotropy convergence was carefully checked; the difference in anisotropy energies calculated using 10,000 and 30,000 k-points was less than 2 percent.